\pdfoutput=1
\documentclass[twocolumn, superscriptaddress, prb]{revtex4-1}
\usepackage{lineno}
\usepackage[separate-uncertainty=true]{siunitx}
\usepackage{graphicx}

\begin{document}

\title{Quantum control of surface acoustic wave phonons}

\date{April 10, 2018}

\author{K. J. Satzinger}
\affiliation{Department of Physics, University of California, Santa Barbara, CA 93106, USA}
\affiliation{Institute for Molecular Engineering, University of Chicago, IL 60637, USA}
\author{Y. P. Zhong}
\affiliation{Institute for Molecular Engineering, University of Chicago, IL 60637, USA}
\author{H.-S. Chang}
\affiliation{Institute for Molecular Engineering, University of Chicago, IL 60637, USA}
\author{G. A. Peairs}
\affiliation{Department of Physics, University of California, Santa Barbara, CA 93106, USA}
\affiliation{Institute for Molecular Engineering, University of Chicago, IL 60637, USA}
\author{A. Bienfait}
\affiliation{Institute for Molecular Engineering, University of Chicago, IL 60637, USA}
\author{Ming-Han Chou}
\affiliation{Institute for Molecular Engineering, University of Chicago, IL 60637, USA}
\affiliation{Department of Physics, University of Chicago, IL 60637, USA}
\author{A. Y. Cleland}
\affiliation{Institute for Molecular Engineering, University of Chicago, IL 60637, USA}
\author{C. R. Conner}
\affiliation{Institute for Molecular Engineering, University of Chicago, IL 60637, USA}
\author{\'E. Dumur}
\affiliation{Institute for Molecular Engineering, University of Chicago, IL 60637, USA}
\affiliation{Institute for Molecular Engineering and Materials Science Division, Argonne National Laboratory, Argonne, IL 60439, USA}
\author{J. Grebel}
\affiliation{Institute for Molecular Engineering, University of Chicago, IL 60637, USA}
\author{I. Gutierrez}
\affiliation{Institute for Molecular Engineering, University of Chicago, IL 60637, USA}
\author{B. H. November}
\affiliation{Institute for Molecular Engineering, University of Chicago, IL 60637, USA}
\author{R. G. Povey}
\affiliation{Institute for Molecular Engineering, University of Chicago, IL 60637, USA}
\affiliation{Department of Physics, University of Chicago, IL 60637, USA}
\author{S. J. Whiteley}
\affiliation{Institute for Molecular Engineering, University of Chicago, IL 60637, USA}
\affiliation{Department of Physics, University of Chicago, IL 60637, USA}
\author{D. D. Awschalom}
\affiliation{Institute for Molecular Engineering, University of Chicago, IL 60637, USA}
\affiliation{Institute for Molecular Engineering and Materials Science Division, Argonne National Laboratory, Argonne, IL 60439, USA}
\author{D. I. Schuster}
\affiliation{Department of Physics, University of Chicago, IL 60637, USA}
\author{A. N. Cleland}
\affiliation{Institute for Molecular Engineering, University of Chicago, IL 60637, USA}
\affiliation{Institute for Molecular Engineering and Materials Science Division, Argonne National Laboratory, Argonne, IL 60439, USA}

\maketitle

\textbf{
The superposition of quantum states is one of the hallmarks of quantum physics, and clear demonstrations of superposition have been achieved in a number of quantum systems. However, mechanical systems have remained a challenge, with only indirect demonstrations of mechanical state superpositions, in spite of the intellectual appeal and technical utility such a capability would bring\cite{Stannigel2010,Kurizki2015}.
This is due in part to the highly linear response of most mechanical systems, making quantum operation difficult, as well as their characteristically low frequencies, making it difficult to reach the quantum ground state\cite{OConnell2010,Teufel2011,Chan2011,Wollman2015,Peterson2016,Chu2017}.
In this work, we demonstrate full quantum control of the mechanical state of a macroscopic mechanical resonator. We strongly couple a surface acoustic wave\cite{Morgan2007} resonator to a superconducting qubit, using the qubit to control and measure quantum states in the mechanical resonator. Most notably, we generate a quantum superposition of the zero and one phonon states and map this and other states using Wigner tomography\cite{Law1996,Banaszek1999,Bertet2002,Haroche2006,Hofheinz2009,Vlastakis2013}.
This precise, programmable quantum control is essential to a range of applications of surface acoustic waves in the quantum limit, including using surface acoustic waves to couple disparate quantum systems\cite{Schuetz2015,Whiteley2018prep}.
}

Linear resonant systems are traditionally challenging to control at the level of single quanta, as they are always in the correspondence limit\cite{Bohr1920}.
The recent advent of engineered quantum devices in the form of qubits has enabled full quantum control over some linear systems, in particular electromagnetic resonators\cite{Hofheinz2009,Vlastakis2013}.
A number of experiments have demonstrated that qubits may provide similar control over mechanical degrees of freedom, including qubits coupled to bulk acoustic modes\cite{OConnell2010,Chu2017}, surface acoustic waves\cite{Gustafsson2014,Manenti2017}, and flexural modes in suspended beams\cite{Arcizet2011,Kolkowitz2012,Yeo2013,Lee2016}.
Of particular note are experiments in which a superconducting qubit is coupled via a piezoelectric material to a microwave-frequency bulk acoustic mode\cite{Cleland2004}, where the ground state can be achieved at moderate cryogenic temperatures, and demonstrations include controlled vacuum Rabi swaps between the qubit and the mechanical mode\cite{OConnell2010,Chu2017}. However, the level of quantum control and measurement has been limited by the difficulty of engineering a single mechanical mode with sufficient coupling and quantum state lifetime.
More advanced operations, such as synthesizing arbitrary acoustic quantum states and measuring those states using Wigner tomography, remain a challenge.
Here we report a significant advance in the level of quantum control of a mechanical device, where we couple a superconducting qubit to a microwave-frequency surface acoustic wave resonance, demonstrating ground-state operation, vacuum Rabi swaps between the qubit and the acoustic mode, and the synthesis of mechanical Fock states as well as a Fock state superposition. We map out the Wigner function for these mechanical states using qubit-based Wigner tomography. We note that a similar achievement has recently been reported with an experiment coupling a superconducting qubit to a bulk acoustic mode\cite{Chu2018prep}.

The device we used for this experiment is shown in Fig.~\ref{fig:device}.
The superconducting qubit is a frequency-tunable planar transmon\cite{Koch2007,Barends2013}, connected to the surface acoustic wave (SAW) device through a tunable inductor network that affords electronic control\cite{Chen2014} of the coupling strength $g$  (see Supplementary Information). 
Qubit control is achieved through two microwave lines for $XY$ and $Z$ control, respectively. 
We measure the qubit state using a dispersively coupled readout resonator (see Supplementary Information). The superconducting qubit is fabricated on a sapphire substrate with standard techniques (see Supplementary Information). 
The SAW resonator is fabricated separately on a lithium niobate substrate, a strong piezoelectric material commonly used for SAW devices\cite{Morgan2007}. 
The SAW resonator comprises an interdigital transducer placed between two Bragg mirrors, designed to support a single surface acoustic wave resonance in the mirror stop band\cite{Morgan2007} (see Supplementary Information). The SAW wavelength $\lambda$ is set by the period of the metal lines that constitute the resonator; here, $\lambda=\SI{1}{\micro m}$, corresponding to a frequency of \SI{4.0}{GHz}. 
At the experiment temperature $T\approx\SI{10}{mK}$, the surface acoustic waves, as well as the qubit, should be in their quantum ground states. 
The electromechanical properties of the SAW resonator are modeled using an equivalent electrical circuit with a complex, frequency-dependent acoustic admittance\cite{Morgan2007} $Y_a(\omega)$ in parallel with an interdigital capacitance $C_t=\SI{0.75}{pF}$. The admittance embeds the complete response of the SAW transducer and the SAW interaction with the mirrors. Lithium niobate's strong electromechanical coupling makes it feasible to strongly couple the SAW resonance to a standard transmon-style qubit (see Supplementary Information).
The separate qubit and SAW resonators chips are connected together in a flip-chip assembly, and coupling between the two chips is achieved using two overlaid planar inductors, one on each chip.
The coupling strength is controlled using an rf squid tunable coupler\cite{Chen2014}, where an externally-controlled flux bias $\Phi_G$ controls the path of the qubit current.

\begin{figure}[t]
	\begin{center}
		\includegraphics[width=3.5in]{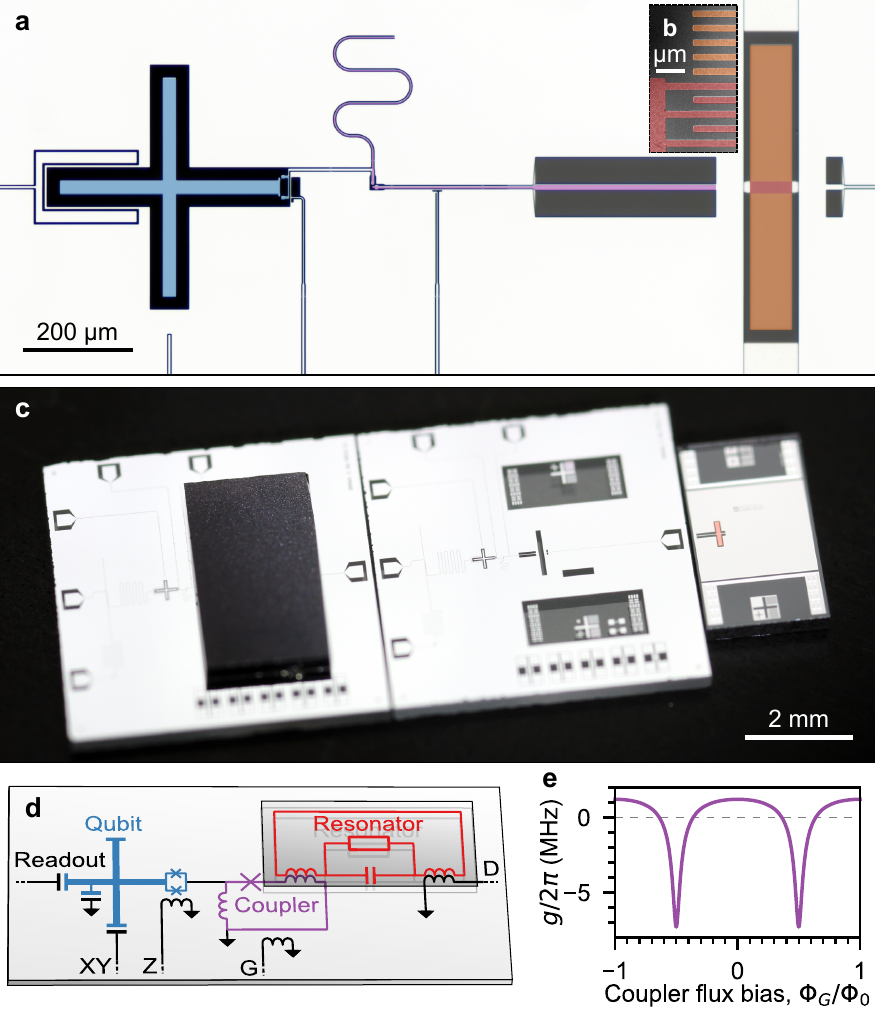}
	\end{center}
\end{figure}
\begin{figure}[t] 
	\caption{
		\label{fig:device}
		{\bf Device description.}
		{\bf a,}
		False-color optical micrograph of a transmon qubit (left, blue) and a SAW resonator (right, red transducer, orange mirrors) which interact via a tunable coupler (center, purple). The device is viewed from below through the transparent sapphire substrate, with the SAW resonator viewed through the sapphire chip on a separate lithium niobate substrate, the two separated by about \SI{7}{\micro m}. 
		{\bf b,}
		Scanning electron micrograph of the SAW resonator with false color on the patterned aluminum film. Red: Upper left corner of the transducer. Orange: Mirror.
		{\bf c,}
		Photograph showing the flip-chip assembly. 
		Right: \SI{2x4}{mm} lithium niobate chip with SAW resonator (red) connected to coupling inductors (horizontal lines).
		Center: \SI{6x6}{mm} sapphire chip with qubit, coupler, and control wiring.
		Left: Flip-chip assembly. The SAW resonator lithium niobate chip (dark rectangle) is inverted, aligned, and affixed to the qubit sapphire chip (see Supplementary Information).
		{\bf d,}
		Schematic circuit diagram, drawn in perspective. Each labeled control line corresponds to an external control or measurement line. Microwave line XY excites the qubit, and line Z controls the qubit frequency. Line G controls the coupler. Line D coherently displaces the resonator state. The qubit, coupler, and control lines are on one plane. The resonator is on a separate chip, represented by the small gray rectangle floating above the qubit plane. The overlaid inductors experience mutual inductive coupling.
		{\bf e,}
		Qubit-resonator coupling $g/2\pi$ calculated for a range of coupler flux bias values $\Phi_G$ using the linear circuit model in {\bf d} with parameters extracted from experiments (see Supplementary Information).
	}
\end{figure}

\begin{figure}
	\begin{center}
		\includegraphics[width=3.5in]{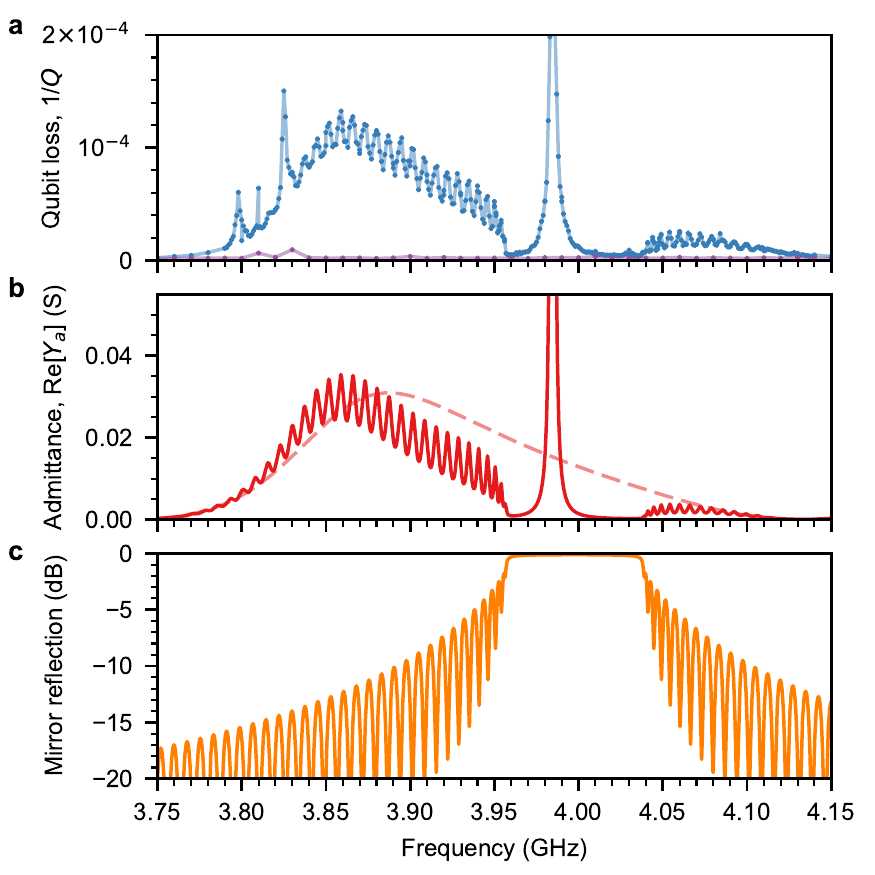}
	\end{center}
	\caption{
		\label{fig:pmatrix}
		{\bf Characterization and modeling of SAW admittance.}
		{\bf a,}
		Measured qubit loss $1/Q$ as a function of qubit frequency $\omega_{ge}/2\pi$. Blue: $|g|/2\pi=\SI{2.3\pm0.1}{MHz}$. Purple: $g$ is minimized.
		{\bf b,}
		Real part of SAW resonator acoustic admittance $\textrm{Re}[Y_a]$, calculated with a numerical model (see Supplementary Information). Red line: Admittance of the full resonator model. The SAW resonance is the large peak at 3.985 GHz. Pink dashed line: Admittance calculated for the transducer alone, without the mirror structure. 
		{\bf c,}
		Magnitude of the model mirror reflection.
	}
\end{figure}

We use the qubit to characterize the coupling and properties of the SAW resonator.
With the coupling off, $g=0$, we observe $T_1\approx\SI{20}{\micro s}$ and $T_{2,\textrm{Ramsey}}\approx\SI{2}{\micro s}$ over the frequency range \SIrange{3.5}{4.5}{GHz} (see Supplementary Information). 
Adjusting $g$ away from zero shortens the qubit lifetime and makes it strongly frequency-dependent,
as the transducer converts electromagnetic energy from the qubit into acoustic waves.
In Fig.~\ref{fig:pmatrix}, we demonstrate this with $|g|/2\pi$ set to \SI{2.3\pm0.1}{MHz}, where acoustic loss is the dominant decay channel for the qubit. We measure the qubit lifetime $T_1$ as a function of qubit frequency $\omega_{ge}/2\pi$ and convert that to a quality factor $Q=\omega_{ge}T_{1}$ and the corresponding loss $1/Q$.
We compare our measurements to a numerical model\cite{Morgan2007} based on the SAW resonator design with parameters fine-tuned to reproduce the frequency response observed in the qubit loss (see Supplementary Information). 
The SAW transducer itself can efficiently emit phonons over a wide range of frequencies, roughly from \SIrange{3.8}{4.1}{GHz}, owing to its small number of finger pairs\cite{Morgan2007} (20 pairs).
The SAW mirror reflects efficiently in the mirror stop band from \SIrange{3.96}{4.04}{GHz}. The resultant interference frustrates the transducer emission except when a resonance condition is met, in this case at the single SAW resonance frequency of $\omega_r/2\pi=\SI{3.985}{GHz}$. The resonator admittance near that resonance can be approximated by an equivalent resonant electrical circuit, constituting the Butterworth van-Dyke model\cite{Morgan2007}. Outside the mirror stop band, the mirror reflection decreases rapidly, and the transducer is free to emit traveling phonons. The qubit sees this as increased loss, especially from \SIrange{3.85}{3.90}{GHz}, where the transducer is most efficient. The ripples in the out-of-band mirror reflection arise from the finite extent of each mirror (500 lines). These features are clearly displayed in the measured qubit loss. The qubit also weakly couples to unidentified resonances near \SI{3.8}{GHz}.
The SAW resonance at \SI{3.985}{GHz} can resonantly and rapidly exchange energy with the qubit.
In subsequent experiments, we avoid unwanted qubit loss by normally keeping the coupling small, only pulsing the coupling on when deliberately interacting with the SAW resonance.

\begin{figure}
	\begin{center}
		\includegraphics[width=3.5in]{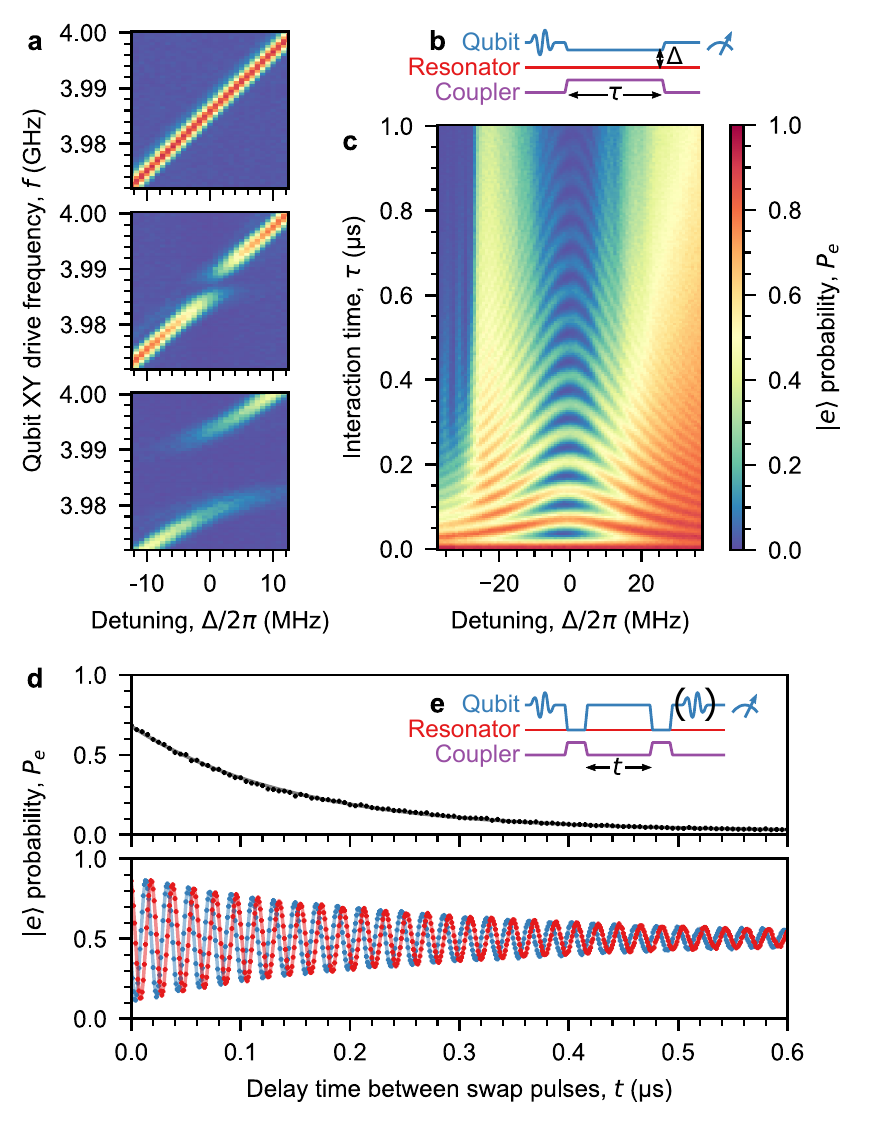}
	\end{center}
	\caption{
		\label{fig:phonon}
		{\bf Qubit interaction with a single mechanical mode.}
		{\bf a,}
		Qubit spectroscopy near the resonator frequency for three different coupler settings. 
		The qubit is biased to frequency $\omega_r+\Delta$ and driven with a \SI{500}{ns} long pulse at frequency $f$; qubit $|e\rangle$ probability $P_e$ is plotted.
		Top: minimum coupling. Middle: moderate coupling, $|g|/2\pi=\SI{2.3\pm0.1}{MHz}$. Bottom: maximum coupling, $|g|/2\pi=\SI{7.3\pm0.1}{MHz}$.
		{\bf b,}
		Rabi-swap pulse sequence. The qubit is excited to $|e\rangle$, and then the qubit is biased to frequency $\omega_r+\Delta$ while the coupling strength is maximized. The qubit and resonator interact for a time $\tau$, and the qubit state is then measured.
		{\bf c,} Probability $P_e$ for the qubit $|e\rangle$ state versus detuning $\Delta$ and interaction time $\tau$. A swap operation is executed by setting the qubit frequency to $\omega_r$ and turning on the coupling for approximately \SI{37}{ns}.
		{\bf d,} Single-phonon experiments using the pulse sequence in {\bf e}. 
		Top: $T_{1r}$ measurement. The qubit is excited to $|e\rangle$, and that excitation is swapped into the resonator. Following a delay time $t$, the state is swapped back into the qubit, and the qubit is measured. 
		Bottom: $T_{2r}$ measurement. The qubit is excited to $|g\rangle-i|e\rangle$, that state is swapped into the resonator, and after a delay time $t$, the state is swapped back into the qubit. We then conduct qubit tomography, a second qubit pulse (blue: $X_{\pi/2}$, red: $Y_{\pi/2}$) followed by qubit measurement (see Supplementary Information).
		The $T_{2r}$ experiment involves generating a quantum superposition of the resonator phonon Fock states $|0\rangle$ and $|1\rangle$. The probabilities oscillate at the idle detuning frequency, $\Delta/2\pi=\SI{53}{MHz}$, exhibiting interference between the resonator state and the qubit tomography pulses.
	}
\end{figure}

We now focus on the interaction between the single SAW resonance and the qubit.
In Fig.~\ref{fig:phonon}a, we illustrate the full range of qubit coupling to the resonance, using spectroscopic measurements of the qubit. We observe a maximum coupling $|g|/2\pi=\SI{7.3\pm0.1}{MHz}$, half the avoided crossing splitting. 
The ratio of the maximum to minimum coupling strength is measured to be at least 300 (see Supplementary Information).
Fig.~\ref{fig:phonon}c shows time-domain Rabi swapping of a single excitation between the qubit and the mechanical mode, representing a photon-phonon exchange each half-oscillation. The number and amplitude of the swaps is primarily limited by the resonator lifetime $T_{1r}$.

We now characterize the quantum state of the resonator. 
We first examine the residual thermal populations in the qubit and resonator excited states, $|e\rangle$ and $|1\rangle$, respectively, using a Rabi population measurement technique\cite{Geerlings2013,Chu2017} (see Supplementary Information). Driven transitions between $|e\rangle$ and the qubit second excited state, $|f\rangle$, are used to quantify the $|e\rangle$ population by measuring the amplitudes of Rabi-like oscillations. 

The experimental results are shown in Fig.~\ref{fig:wigner}a. Performing the experiment on the qubit alone, we observe an excited state population of $0.0169\pm0.0002$. To assess the resonator thermal population, we first execute a swap operation, and then we conduct the experiment again. The swap exchanges the small excited state populations in the resonator and the qubit. In this case, we observe an excited state population of $0.0049\pm0.0002$, which we interpret as an upper bound on the resonator excited state population\cite{Chu2017}.

\begin{figure}[h]
	\begin{center}
		\includegraphics[width=3.5in]{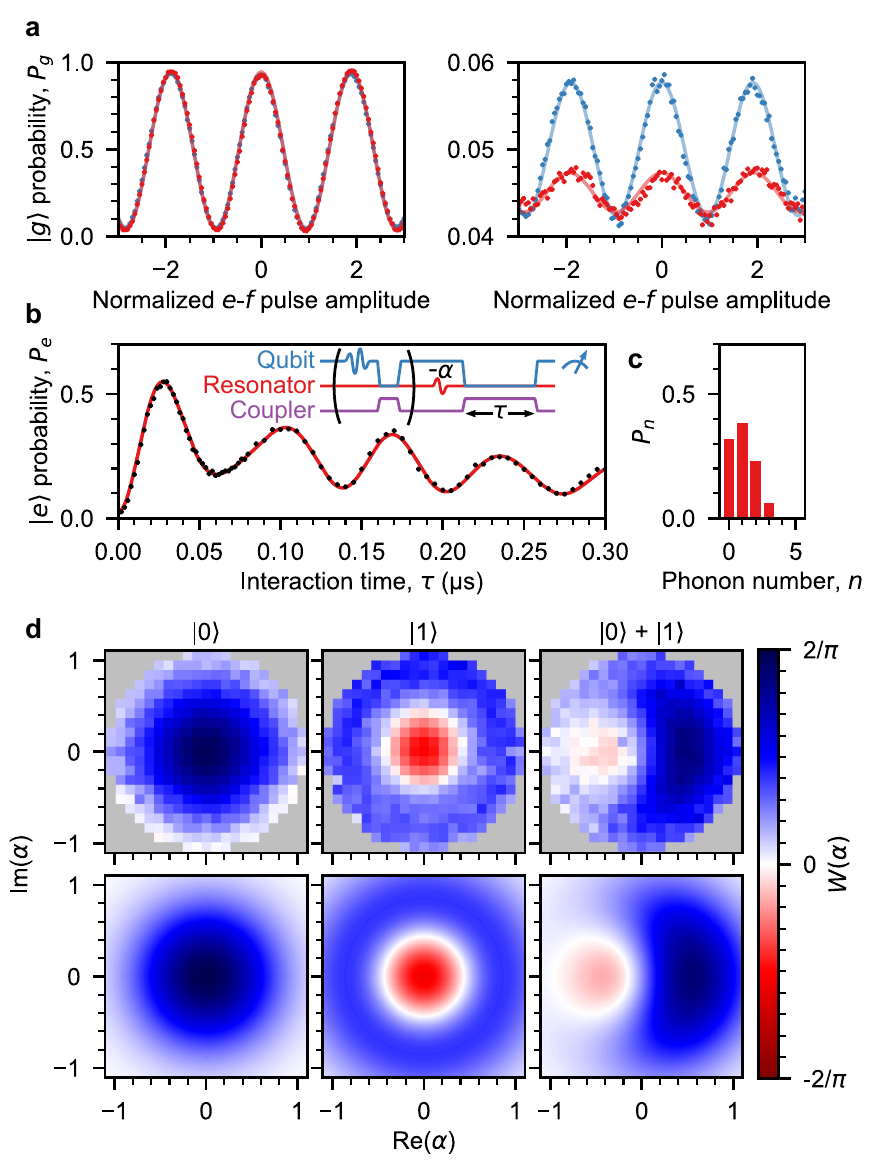}
	\end{center}
\end{figure}
\begin{figure}[t] 
	\caption{
		\label{fig:wigner}
		{\bf Resonator state characterization.}
		{\bf a,}
		Rabi population measurement\cite{Geerlings2013} to determine the steady-state qubit $|e\rangle$  population (see Supplementary Information). 
		The sequences to probe ground state (left) and excited state (right) populations are performed on the equilibrium qubit state (blue) and the qubit state following a swap operation with the resonator (red)\cite{Chu2017}.
		Left: Large-amplitude oscillations showing near-unity initial ground state populations.
		Right: Oscillations showing small initial $|e\rangle$ populations.
		We calculate the excited state populations from the amplitudes of these oscillations.
		Negative values on the horizontal axis correspond to $e-f$ pulses with relative phase of $\pi$ radians.
		$e-f$ pulse amplitude is normalized to the amplitude which approximately swaps $|e\rangle$ and $|f\rangle$, measured separately.
		{\bf b,}
		Example Wigner tomography experiment showing the qubit evolution as it interacts with a displaced resonator $|1\rangle$ state (black points).
		Inset: 
		Mechanical state synthesis and Wigner tomography pulse sequence. If needed, the qubit is excited to the desired state, which is then swapped into the resonator. To determine the Wigner function $W(\alpha)$, the resonator state is displaced with coherent amplitude $-\alpha$. The qubit interacts with the displaced resonator state for a time $\tau$ before it is measured, allowing the phonon number distribution of the displaced state to be determined.
		{\bf c,}
		Example phonon number distribution $P_n$ resulting from a fit to the experiment in {\bf b} (red line). $W(\alpha)$ is then calculated from the fitted probability distribution. Statistical uncertainty in each probability is approximately 0.004.
		{\bf d,}
		Wigner functions $W(\alpha)$ of SAW resonator quantum states. Top: Experimental results.
		The $|0\rangle+|1\rangle$ Wigner function is rotated \ang{90} to compensate for relative phase accumulation during the pulse sequence\cite{Hofheinz2009}.
		Bottom: Prediction of the numerical model.
		The experimental fidelities are (left to right): $0.985\pm0.005$, $0.858\pm0.007$, and $0.945\pm0.006$. The model predicts similar fidelities: 0.998 (limited by thermal occupation), 0.879 (limited by $T_{1r}$), and 0.962 (limited by $T_{1r}$), respectively.
	}
\end{figure}

We characterize the single-phonon properties of the resonator in Fig.~\ref{fig:phonon}d. We prepare a quantum state in the qubit, swap that state into the resonator, wait a delay time $t$, swap the state back into the qubit, and measure the qubit. The decay of the phonon is consistent with energy lifetime $T_{1r}=\SI{148\pm 1}{ns}$ and dephasing time $T_{2r}=\SI{293\pm 1}{ns}$, where the ratio $T_{2r}/T_{1r}\approx 2$ is consistent with little to no additional phase decoherence, as expected for a harmonic oscillator. 

The level of control achievable in this experiment allows us to controllably generate the resonator states $|0\rangle$, $|1\rangle$, $|0\rangle+|1\rangle$, and $|2\rangle$ (see Supplementary Information).
We use Wigner tomography to determine the fidelities of these quantum states\cite{Hofheinz2009} (see Supplementary Information), examining the three lowest-energy states in detail. Following state preparation, we measure the Wigner function $W(\alpha)$ of the resonator by using the qubit to measure the parity of the resonator states at different complex displacements $\alpha$ in resonator phase space (see Supplementary Information).
The required displacements $\alpha$ are created by driving the resonator with a resonant Gaussian microwave pulse applied to a control line (see Fig.~\ref{fig:device}d). During the pulse, the coupling is turned off, and the qubit is detuned above the resonator by $\Delta/2\pi=\SI{400}{MHz}$.

With the qubit initially in its ground state $|g\rangle$, we allow the qubit and resonator to interact for a time $\tau$, then measure the qubit. The qubit state as a function of delay $\tau$ contains information about the displaced resonator state. We repeat the experiment for many values of $\alpha$. The results are displayed in Fig.~\ref{fig:wigner}d, along with the prediction of the numerical model using the same pulse sequence. We then convert each experimental $W(\alpha)$ into a density matrix $\rho$. 
From the density matrices, we calculate the quantum state fidelities (see Supplementary Information); for the resonator $|0\rangle+|1\rangle$ state, we find a fidelity of $0.945\pm0.006$.

In conclusion, we demonstrate high-fidelity, on-demand synthesis of quantum states in a macroscopic mechanical resonator. 
This demonstration involves a hybrid architecture incorporating a high-performance qubit with strong tunable coupling to  surface acoustic waves. 
This scalable platform holds promise for future quantum acoustics experiments coupling stationary qubits to ``flying" qubits based on phonons, and possibly for coupling together other diverse quantum systems such as spins in semiconductors.

\subsection*{Acknowledgements}
We thank P. J. Duda, A. Dunsworth, and D. Sank for helpful discussions. Devices and experiments were supported by the Air Force Office of Scientific Research and the Army Research Laboratory, and material for this work was supported by the Department of Energy (DOE). K.J.S. and S.J.W. were supported by NSF GRFP (NSF DGE-1144085), \'E.D. was supported by LDRD funds from Argonne National Laboratory, A.N.C. and D.D.A. were supported by the DOE, Office of Basic Energy Sciences, and D.I.S. acknowledges support from the David and Lucile Packard Foundation. This work was partially supported by the UChicago MRSEC (NSF DMR-1420709) and made use of the Pritzker Nanofabrication Facility, which receives support from SHyNE, a node of the National Science Foundation's National Nanotechnology Coordinated Infrastructure (NSF NNCI-1542205).

\subsection*{Author Contributions}
K.J.S. designed and fabricated the devices. K.J.S., H.S.C., J.G., A.Y.C., and S.J.W. developed the fabrication processes. G.A.P., \'E.D., and A.N.C. contributed to device design. K.J.S. performed the experiments and analyzed the data with assistance from Y.P.Z., A.B., and \'E.D. Significant assistance was provided by I.G. and B.H.N. A.N.C., D.I.S. and D.D.A. advised on all efforts. All authors contributed to discussions and production of the manuscript.

\subsection*{Author Information}
The authors declare no competing financial interests. Correspondence and requests for materials should be addressed to A. N. Cleland (anc@uchicago.edu).

\clearpage

\bibliography
\bibliographystyle{naturemag}

\clearpage
\begin{center}
	\textbf{\large Supplementary Information for ``Quantum control of surface acoustic wave phonons"}
\end{center}
\setcounter{equation}{0}
\setcounter{figure}{0}
\setcounter{table}{0}
\makeatletter
\renewcommand{\theequation}{S\arabic{equation}}
\renewcommand{\thefigure}{S\arabic{figure}}
\renewcommand{\thetable}{S\arabic{table}}

\section{Device details}
\subsection{Qubit chip}
We fabricate the qubit chip on a sapphire substrate using standard methods adapted from refs.~\onlinecite{Kelly2015thesis,Dunsworth2017} with six steps.

(I) Al deposition, \SI{100}{nm}.

(II) Al etch defining qubit capacitor, readout, and control circuitry. Photolithographically patterned and etched with BCl$_3$/Cl$_2$/Ar inductively coupled plasma.

(III) Crossover SiO$_2$ deposition, \SI{200}{nm}, patterned with liftoff.

(IV) Crossover Al deposition, \SI{230}{nm}, preceded by in situ Ar ion mill, patterned with liftoff.

(V) Josephson junction deposition using Dolan bridge shadow evaporation and liftoff, using a PMMA/MAA bilayer and electron beam lithography. Not preceded by Ar ion mill.

(VI) Bandage Al deposition\cite{Dunsworth2017}, preceded by in situ Ar ion mill, patterned with liftoff. This step establishes galvanic connections between the aluminum from (I) and (V).

We use electron beam evaporation to deposit each film. We use photolithography with \SI{0.9}{\micro m} i-line photoresist (AZ MiR 703) for (II), (III), (IV), (VI). Each liftoff step is in N-methyl-2-pyrrolidone at \SI{80}{\celsius}.

The qubit and coupler design is based on refs.~\onlinecite{Chen2014,Neill2016,Geller2015}. For dispersive readout, the qubit is capacitively coupled to a \SI{5.4}{GHz} quarter-wave coplanar waveguide resonator which is inductively coupled to a quarter-wave bandpass filter\cite{Jeffrey2014,Kelly2015}. 

\subsection{SAW resonator chip}
We fabricate the SAW resonator chip on a lithium niobate substrate. First, we pattern an Al film to define the transducer, mirrors, coupling inductors, and ground plane. We create the pattern with a PMMA 950K/PMMA 495K bilayer and electron beam lithography, followed by electron beam evaporation of Al (\SI{25}{nm}) and liftoff in acetone. Second, we pattern SU-8 epoxy spacers (\SI{6.5}{\micro m} thick) on the periphery of the chip. These spacers determine the separation between the chips during flip-chip assembly.

The SAW transducer and mirror layout is designed to give a single mode with a continuous grating structure\cite{Ebata1988,Hashimoto2004,Morgan2007}.

\subsection{Flip-chip assembly}
We assemble completed qubit and SAW resonator chips using a standard manual mask aligner (Karl-Suss MJB4). A machined acrylic plate transfers the ``mask" vacuum to the transparent, double-side-polished sapphire chip, which is suspended face-down. We manually apply \SI{\approx 10}{nL} of glue to the periphery of the SAW resonator chip. The glue choice is not critical; we use nLOF 2070 photoresist. The SAW resonator chip is placed on the sample chuck of the aligner, and we align the chips and bring them into contact. Our glue requires a bake to solidify; we leave the assembly clamped together in the aligner and heat the assembly with a hot air gun (estimated chip temperature \SI{60}{\celsius}). Typical alignment error is \SI{<2}{\micro m}.

\section{Models}

\subsection{SAW resonator, electromechanical model}
In Fig.~\ref{fig:pmatrix}, we show results from a numerical model of the SAW resonator. We use a standard 1D electromechanical model, the $P$-matrix\cite{Morgan2007,Hashimoto2004}. The full model is composed of a transducer model and a mirror model, both using the coupling-of-modes method. We begin with the lithographically-determined device parameters and standard material parameters from ref.~\onlinecite{Morgan2007}. We tune the mirrors' effective wave speed $v_m$ and amplitude reflection per line $r_m$ to reproduce the apparent stop band observed in Fig.~\ref{fig:pmatrix}a. We tune the transducer's speed $v_t$ and reflection $r_t$ to place the resonance at \SI{3.985}{GHz} and reproduce the apparent asymmetric transducer response. We introduce uniform propagation loss $\eta$ in both the transducer and mirror and adjust the loss so that the quality factor $Q$ of an approximating series $RLC$ circuit fitted to the peak in the model admittance $Y_a(\omega)$ is consistent with the $T_{1r}$ measurement in Fig.~\ref{fig:phonon}d. These are the values used in Fig.~\ref{fig:pmatrix}: $v_m=\SI{4027.0}{m/s}$, $v_t=\SI{4012.5}{m/s}$, $r_m=-0.032i$, $r_t=-0.015i$, and $\eta=\SI{851}{Np/m}$. The reported speed for a nonmetallized surface at room temperature is \SI{3979}{m/s}. The mirror and transducer parameters are expected to differ; the metal lines in the mirror are electrically floating, which gives stronger reflectivity. These values are consistent with cryogenic measurements of similar SAW resonators using a vector network analyzer. We also measure multi-mode SAW resonators, incorporating nonmetallized surface between the transducer and mirrors; the propagation loss $\eta$ in the nonmetallized surface appears to be about one tenth the loss in the transducer and mirrors.

The model series $RLC$ circuit gives an equivalent $C_s=\SI{12.10}{fF}$, $L_s=\SI{131.8}{nH}$, and $R_s=\SI{0.890}{\ohm}$.  $C_s/C_t$ is roughly proportional to the piezoelectric coupling strength\cite{Morgan2007}. The large piezoelectric coupling strength of lithium niobate (45 times that of quartz\cite{Morgan2007}) gives a series resonance with a relatively small characteristic impedance $Z=\sqrt{L_s/C_s}$, enhancing the inductive coupling $g$ to the qubit circuit.

\subsection{Linear circuit model}
The qubit chip and the SAW resonator chip are brought together in a flip-chip assembly, where two planar inductors $L\approx\SI{0.3}{nH}$ are brought into close proximity, one on each chip. They experience a mutual inductance $M\approx0.4L$. Qubit current passing through $L$ induces a current in the opposing $L$ on the resonator chip, enabling the qubit to drive the SAW transducer. We employ this mutual inductance in a modified version of previous rf squid tunable couplers\cite{Chen2014,Geller2015}. The coupler has a single junction with unbiased inductance $L_{cj0}=\SI{1.0}{nH}$ whose phase $\delta$, and hence inductance $L_{cj}=L_{cj0}/\cos(\delta)$, is controlled with an external flux bias $\Phi_G$. This tunable inductance creates a current divider; the coupling can be turned off by tuning $L_{cj}$ to be very large, and $|g|$ is maximized when $\Phi_G=0.5\Phi_0$ ($\delta=\pi$, $L_{cj}=-L_{cj0}$). The total inductance of the network is approximately $L$ (the inductance of the planar coupling inductors), much smaller than the qubit equivalent inductance $L_q$ (typically \SI{\approx10}{nH}), so the coupling circuit attaches to the qubit at a low voltage node.

\begin{figure}
	\begin{center}
		\includegraphics[width=3.5in]{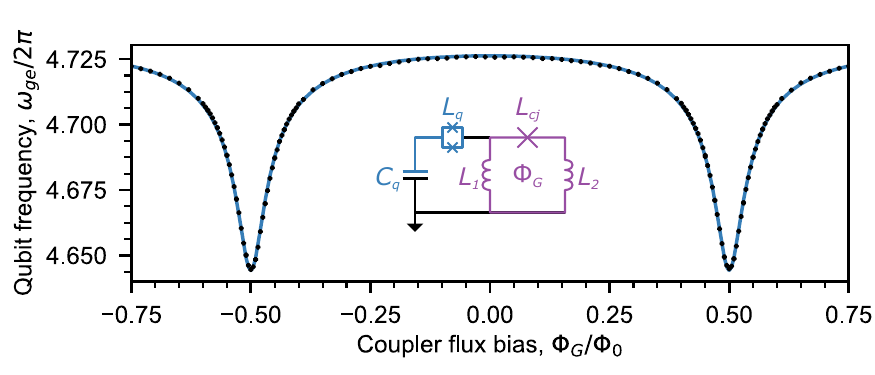}
	\end{center}
	\caption{
		\label{fig:s_qubitfreq}
		{\bf The effect of the coupler on the qubit frequency.}
		Black points: Spectroscopically-determined qubit frequency $\omega_{ge}/2\pi$ as a function of coupler flux bias $\Phi_G$. Blue line: Qubit frequency from a linear circuit model fitted to the experimental data. Inset: Circuit model of the qubit and coupler.
	}
\end{figure}

To determine the qubit and coupler parameters, we use a spectroscopic measurement of the qubit frequency $\omega_{ge}$ as we vary the coupler flux bias $\Phi_G$. We fit the response to a linear circuit model of the qubit and coupler. We set the qubit flux bias $\Phi_Z=0$, and we compensate for linear crosstalk between the two flux biases.
This is shown in Fig.~\ref{fig:s_qubitfreq}. We fix the unbiased coupler junction inductance $L_{cj0}=\SI{1.0}{nH}$ based on room temperature and cryogenic DC measurements of test junctions. The other parameters are fitted: $C_q=\SI{110}{fF}$, $L_q=\SI{10.1}{nH}$, $L_1=\SI{0.303}{nH}$, and $L_2=\SI{0.403}{nH}$. We construct a full linear circuit model following Fig.~\ref{fig:device}d using these parameters together with the series $RLC$ of the SAW resonance, which we use to calculate the coupling $g$ in Fig.~\ref{fig:device}e. We adjusted the mutual inductance $M$ between the coupler and the resonator's coupling inductor to reproduce the experimental coupling; $M=\SI{0.13}{nH}$.

\subsection{Coupled quantum system, master equation model}
We model the quantum behavior of the coupled qubit and SAW resonance using the python package QuTiP\cite{Johansson2012}. The model consists of a 2-level qubit (lowering operator $\sigma_-$) with $T_1$ decay and $T_\phi=((T_{2,\textrm{Ramsey}})^{-1} - (2T_1)^{-1})^{-1}$ dephasing coupled to a 10-level harmonic oscillator (lowering operator $a$) with $T_1$ decay under the Jaynes-Cummings Hamiltonian. We adopt the resonator rotating frame, giving the Hamiltonian\cite{Hofheinz2009} 
\begin{equation}
H=\hbar\Delta\sigma_+\sigma_- + \hbar g(\sigma_+a+\sigma_-a^\dagger). 
\end{equation}
To simulate the time evolution of the composite density matrix $\rho$, we numerically integrate the Lindblad master equation, 
\begin{equation}
\frac{d\rho}{dt}=-\frac{i}{\hbar}[H,\rho]+\sum_n\left(c_n\rho c_n^\dagger-\frac{1}{2}\left\{c_n^\dagger c_n,\rho\right\}\right),
\end{equation}
subject to collapse operators 
$c_n=\sigma_-/\sqrt{T_1}, \sigma_z/\sqrt{2T_\phi}, a/\sqrt{T_{1r}}$.
We determine these parameters by fitting the model to standard qubit measurements and the experiments depicted in Fig.~\ref{fig:phonon}. We use the thermal populations from Fig.~\ref{fig:wigner}a to set the initial states. When simulating measurements of qubit $|e\rangle$ probability $P_e$, we multiply the raw simulated value by a factor of 0.97 to account for readout visibility.

\section{Experiments}
\subsection{Qubit loss}
\begin{figure}
	\begin{center}
		\includegraphics[width=3.5in]{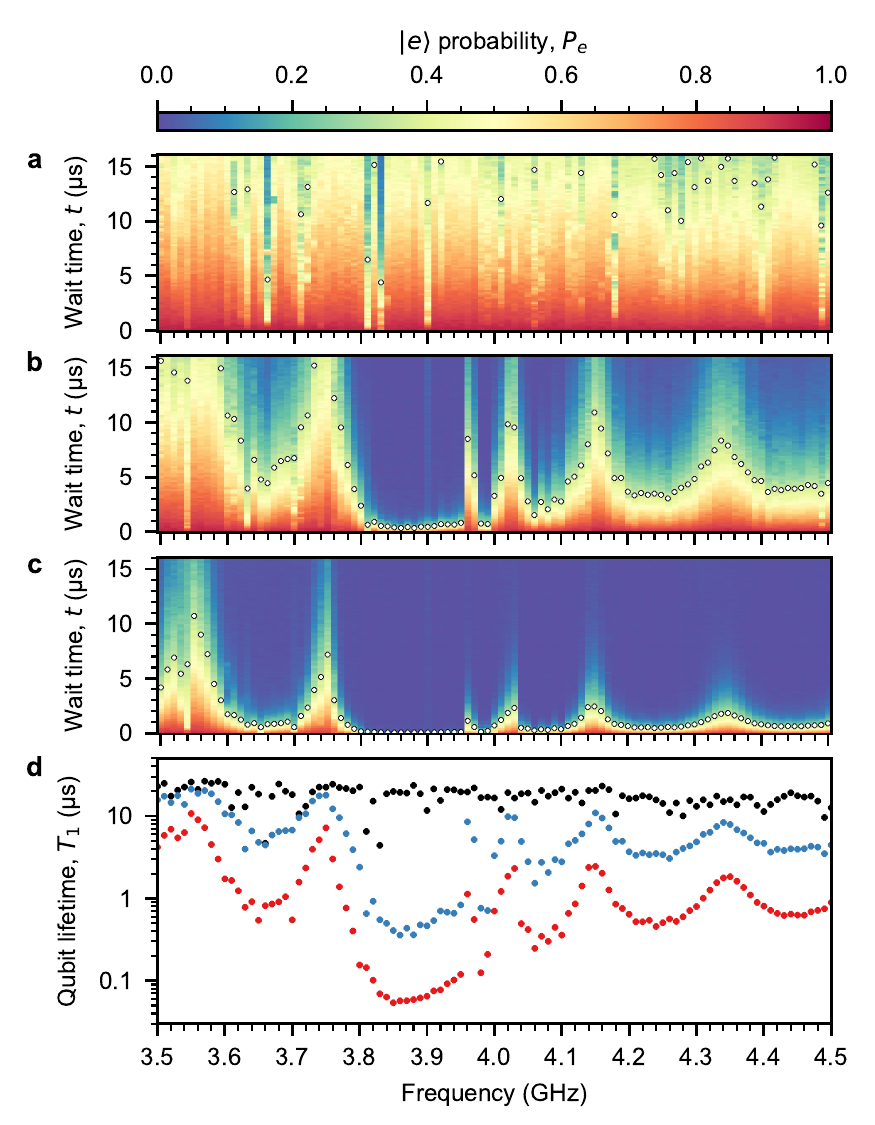}
	\end{center}
	\caption{
		\label{fig:s_t1scan}
		{\bf Qubit $T_1$ scans.}
		{\bf a-c,} Time-domain qubit decay measurements with qubit $|e\rangle$ probability $P_e$ plotted over wait time $t$ for a range of qubit frequencies. Black circles are fitted $T_1$ values. In {\bf a}, many $T_1$ fits were \SI{\approx 20}{\micro s}, above the plot range. The coupler flux $\Phi_G$ is different for each scan: {\bf a} $0.35\Phi_0$, {\bf b} $0.44\Phi_0$, and {\bf c} $0.5 \Phi_0$.
		{\bf d} Fitted $T_1$ values plotted on a logarithmic scale, colored {\bf a} black, {\bf b} blue, and {\bf c} red.
	}
\end{figure}

We measured the qubit lifetime $T_1$ over the frequency range \SIrange{3.5}{4.5}{GHz} for the three coupling conditions used in Fig.~\ref{fig:phonon}a. The results are plotted in Fig.~\ref{fig:s_t1scan}. The coupling is minimized in Fig.~\ref{fig:s_t1scan}a, and the qubit has a fairly consistent lifetime $T_1\approx\SI{20}{\micro s}$ with lower values observed at some frequencies. In Fig.~\ref{fig:s_t1scan}b and c, the coupling is increased, and the qubit exhibits strongly frequency-dependent loss. In Fig.~\ref{fig:pmatrix}, this frequency dependence is explained; Fig.~\ref{fig:pmatrix}a is from a finer scan under the conditions of Fig.~\ref{fig:s_t1scan}b. Fig.~\ref{fig:phonon}c is a finer scan under the conditions of Fig.~\ref{fig:s_t1scan}c near \SI{3.985}{GHz}. We also performed Ramsey experiments under these three coupling conditions, at \SI{3.50}{GHz}, \SI{4.00}{GHz}, and \SI{4.50}{GHz}. In each case, $T_{2,\textrm{Ramsey}}\approx$ \SIrange{1}{3}{\micro s}.

It is difficult to use the resonance to measure the coupling $g$ when it is smaller than the resonator loss rate. To obtain a lower bound on the coupling on/off ratio, we instead look at the behavior at \SI{3.850}{GHz}, where the transducer efficiently launches traveling phonons. This is the qubit's dominant loss mechanism at \SI{3.850}{GHz} in Fig.~\ref{fig:s_t1scan}b and c. At that frequency, with the coupling minimized, $T_1=\SI{19.8}{\micro s}$, and with the coupling maximized, $T_1=\SI{54}{ns}$. Their ratio, 366, is a lower bound on the on/off ratio.

\subsection{Rabi population measurement}
We use the Rabi population measurement method\cite{Geerlings2013} to determine the steady-state populations of the excited states of the qubit and resonator.
The qubit and resonator have small thermal $|e\rangle$ and $|1\rangle$ populations. We determine those populations by driving $|e\rangle-|f\rangle$ qubit transitions; the qubit second excited state $|f\rangle$ thermal population is assumed to be negligible, as validated by these experiments. 

\begin{figure}
	\begin{center}
		\includegraphics[width=3.5in]{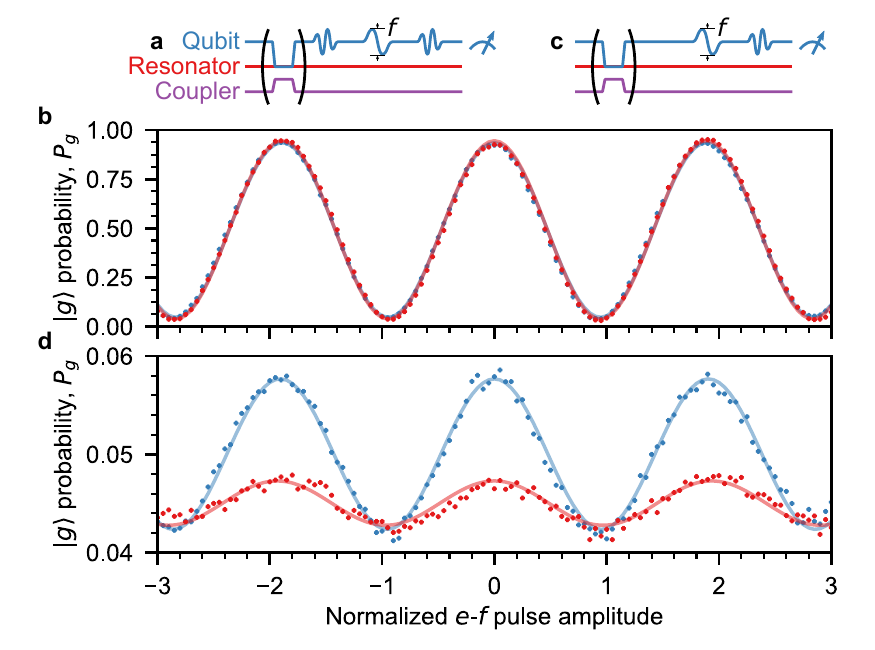}
	\end{center}
	\caption{
		\label{fig:s_thermometry}
		{\bf Qubit and resonator thermometry.} Also see Fig.~\ref{fig:wigner}.
		{\bf a,} Rabi population measurement pulse sequence for the ground state measurement. Following an optional swap operation, the qubit is excited with an $X_\pi$ pulse before being driven by a pulse at $\omega_{ef}$ with variable amplitude and finally another $X_\pi$ pulse.
		{\bf b,} Ground state measurement results. 
		{\bf c,} Pulse sequence for the excited state measurement, the same as {\bf a} without the initial $X_\pi$ pulse.
		{\bf d,} Excited state measurement results. Blue: Qubit alone (no swap). Red: Swap. Points are from measurements; lines are cosine fits.
	}
\end{figure}

In Fig.~\ref{fig:s_thermometry}a-b, the qubit is excited with an $X_\pi$ pulse prior to the $\omega_{ef}$ pulse. This places most of the population in $|e\rangle$, and that large population is coherently driven between $|e\rangle$ and $|f\rangle$ by the $\omega_{ef}$ pulse, resulting in a superposition of $|e\rangle$ and $|f\rangle$. Finally, a second $X_\pi$ pulse swaps whatever is in $|e\rangle$ with what is in $|g\rangle$, and we measure $|g\rangle$ probability $P_g$. This probability exhibits Rabi-like oscillations with the $\omega_{ef}$ pulse amplitude. The near-unity peak-to-peak amplitude of those oscillations $A_g$ reflects the large initial $|g\rangle$ population.

In Fig.~\ref{fig:s_thermometry}c-d, the same experiment is carried out without the initial $X_\pi$ pulse. Thus the $\omega_{ef}$ pulse drives oscillations of the small thermal population between $|e\rangle$ and $|f\rangle$. Again, the ending $X_\pi$ pulse swaps the $|e\rangle$ and $|g\rangle$ populations, where we measure it. The small amplitude of those oscillations $A_e$ reflects the small initial $|e\rangle$ population. We calculate the initial $|e\rangle$ population $P_e=A_e/(A_e+A_g)$, obtaining $0.0169\pm0.0002$ for the qubit.

Following ref.~\onlinecite{Chu2017}, we use this technique to establish a lower bound on the SAW resonator ground state probability. Since the thermal populations of the qubit and resonator are both small, those populations are exchanged using a qubit-resonator swap. We repeat the experiment immediately preceded by a swap, obtaining $P_e=0.0049\pm0.0002$, an upper bound on the SAW resonator initial $|1\rangle$ population. This implies the SAW resonator ground state probability is at least 99.5\%.

\subsection{Qubit tomography}
The $T_{2r}$ measurement in Fig.~\ref{fig:phonon}d uses qubit tomography to measure the qubit along the $X$ and $Y$ axes. The standard qubit $P_e$ measurement follows a tomography pulse. We repeat the experiment using each of these tomography pulses: $X_{\pi/2}$, $X_{-\pi/2}$, $Y_{\pi/2}$, $Y_{-\pi/2}$, $X_\pi$, $X_{-\pi}$, $Y_\pi$, $Y_{-\pi}$, and no pulse. The negative phase pulses ensure a symmetric measurement. For example, the qubit measurement along the $Y$ direction is $[P(X_{-\pi/2}) + (1-P(X_{\pi/2}))]/2$. Put that way, in Fig.~\ref{fig:phonon}d, blue is measurement along $-Y$, and red is measurement along $X$. The tomography allows construction of a Bloch vector representing the qubit state with entries $(\langle\sigma_X\rangle,\langle\sigma_Y\rangle,\langle\sigma_Z\rangle)$, where $\langle\sigma_i\rangle=2P(i)-1$ is the expectation value of the Pauli operator $i$, and $P(i)$ is the measured probability of the qubit along direction $i$.

\begin{figure}[h]
	\begin{center}
		\includegraphics[width=7.0in]{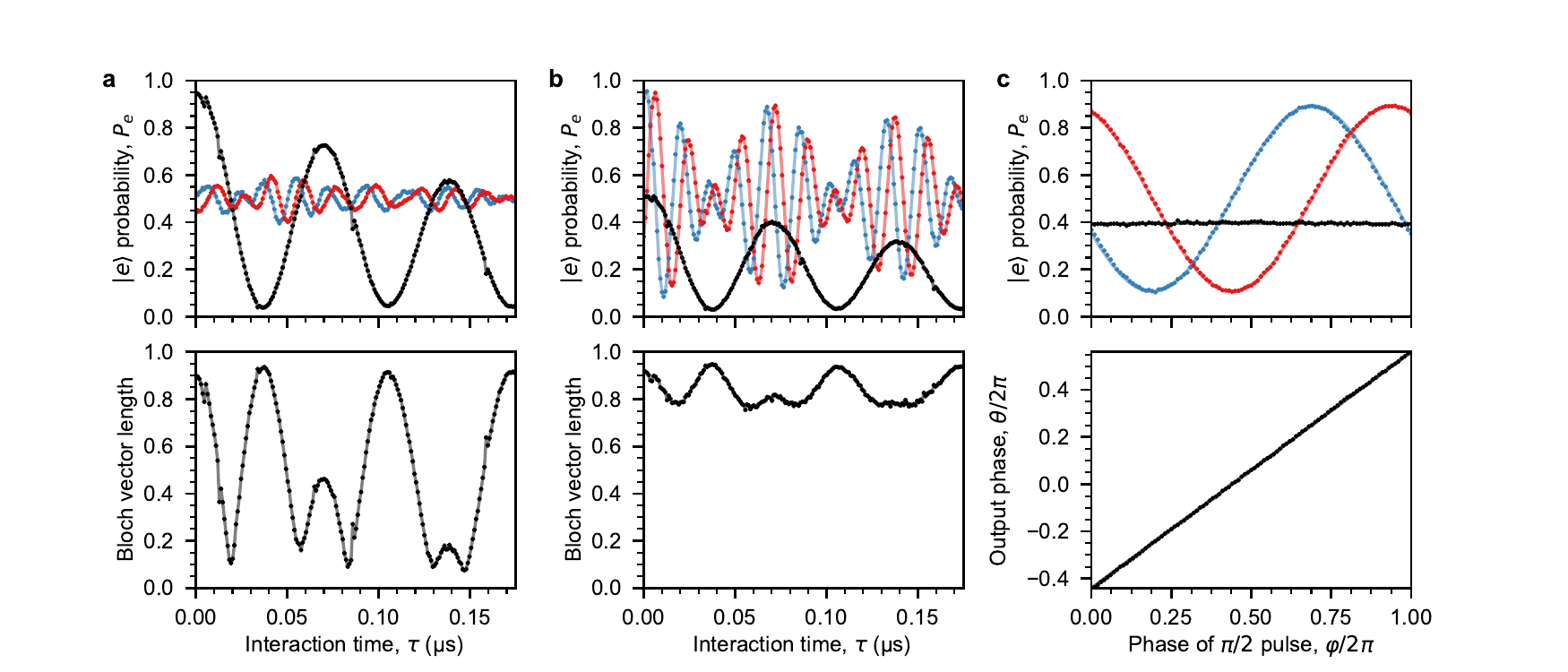}
	\end{center}
	\caption{
		\label{fig:s_qtomo}
		{\bf Qubit tomography and interaction with the SAW resonator.} 
		Top row: Qubit tomography results. Black, blue, and red show measurement along $Z$, $-Y$, and $X$, respectively.
		{\bf a,} Resonant Rabi swapping, as in Fig.~\ref{fig:phonon}c at $\Delta=0$. The qubit starts in $|e\rangle$. Top: Qubit tomography. Bottom: Bloch vector length calculated from tomography.
		{\bf b,} Resonant Rabi swapping with the qubit starting in $|g\rangle-i|e\rangle$.
		{\bf c,} Measurement of the phase of superposition states after swapping in and out of the resonator. Top: Qubit tomography of the final state with constant and sinusoidal fits. Bottom: Final state phase $\theta$, measured from the $X$ axis, with unity-slope linear fit.
	}
\end{figure}

We conduct additional experiments with qubit tomography to study the interaction between the qubit and SAW resonator. 
In Fig.~\ref{fig:s_qtomo}a, we show Rabi swapping between $|e,0\rangle$ and $|g,1\rangle$. Ideally, the $X$ and $Y$ measurements would be at $P_e=0.5$; imperfections in state preparation or the swap pulses introduce errors resulting in small oscillations about $P_e=0.5$ at the idle detuning frequency. We also plot the length of the qubit Bloch vector, which becomes small halfway through a swap, when the qubit is near a uniform statistical mixture. This is because we only measure the qubit, while some of the energy is left unmeasured in the resonator. The Bloch vector recovers on each oscillation, suggesting entanglement between the qubit and resonator.
In Fig.~\ref{fig:s_qtomo}b, we show a similar experiment with the qubit starting in the superposition $|g\rangle-i|e\rangle$. Following one swap, the qubit is near the ground state, and as the state swaps back into the qubit, we observe large $X$ and $Y$ oscillations, showing that the superposition persists. In Fig.~\ref{fig:s_qtomo}c, we demonstrate control of the phase of the superposition of the SAW resonator. The qubit starts in $|g\rangle-ie^{i\phi}|e\rangle$. The state is swapped into the resonator, we wait 5 ns, and then the state is swapped back to the qubit. We then measure the final state phase $\theta$; it equals $\phi$ plus an offset determined by the relative phase accumulated during the sequence.

\onecolumngrid

\twocolumngrid

\subsection{Wigner tomography}
We conduct Wigner tomography following the method in ref.~\onlinecite{Hofheinz2009}. In the experiment, we measure the qubit-resonator interaction for different displacements $\alpha$ of the resonator state in resonator phase space, where $\alpha$ is complex valued. 
For each $\alpha$ value and each resonator state, we record the qubit state after different interaction times with the displaced resonator state.
We then use the master equation model to deduce the diagonal elements of the displaced resonator state's density matrix, $\rho_{nn}^\prime(\alpha)$, which constitute a probability distribution $P_n(\alpha)$ (here, $n=0, 1, \ldots, 9$). 
We measure the qubit state prior to the evolution to establish an initial mixed state in the qubit (typical $P_e\approx0.03$). We use a cost function which takes a candidate distribution of resonator populations $P_n$, generates the evolution of $P_e$ predicted by the model, and returns the summed squared error between $P_e$ as predicted by the model and $P_e$ as measured in the experiment. We numerically minimize this function to arrive at a fitted $P_n(\alpha)$. We assess the uncertainty in $P_n$ by numerically calculating the second derivative of the error with respect to each probability in a distribution.

We use all of the $P_n(\alpha)$ values to determine the density matrix $\rho$ of each state. We fit to $4\times 4$ density matrices using 15 real parameters (expanding in generalized Gell-Mann matrices\cite{bertlmann2008}). We convert the $4\times 4$ matrices into $10\times 10$ to accommodate the displacement operations. We minimize a cost function which takes 15 real values, converts them into a candidate density matrix $\rho$, displaces $\rho$ by each experimental $\alpha$ value, and compares the diagonal elements of the displaced $\rho$ to the experimental $P_n(\alpha)$ values. In this case, we directly obtain variance-covariance matrices which establish the uncertainties in each parameter (typically $\approx0.008$, which translates to a similar error in each element of $\rho$). The fitted density matrices typically have small negative eigenvalues due to noise ($\approx-0.02$); we truncate these to zero and renormalize the density matrices. Finally, we compute the fidelities by comparing to the ideal pure states $|\psi\rangle$, $\sqrt{\langle\psi|\rho|\psi\rangle}$. We estimate the error in the fidelity using Monte Carlo error propagation; the dominant source is from fitting $\rho$, not $P_n(\alpha)$.

For the resonator displacement, we apply a Gaussian microwave pulse to the D control line (see Fig.~\ref{fig:device}d). The pulse has a full-width half-max of \SI{3}{ns} and is centered at the SAW resonance frequency. The amplitude of the pulse is proportional to the displacement amplitude $\alpha$. We calibrate the proportionality with a separate experiment, where we excite the resonator with a displacement pulse, perform a swap, and measure the qubit. We repeat the experiment for many pulse amplitudes and then fit the experiment to our master equation model with one parameter; that parameter is the calibration to convert between experimental amplitudes and $\alpha$ values.

\subsection{Additional experiments}
\begin{figure}
	\begin{center}
		\includegraphics[width=3.5in]{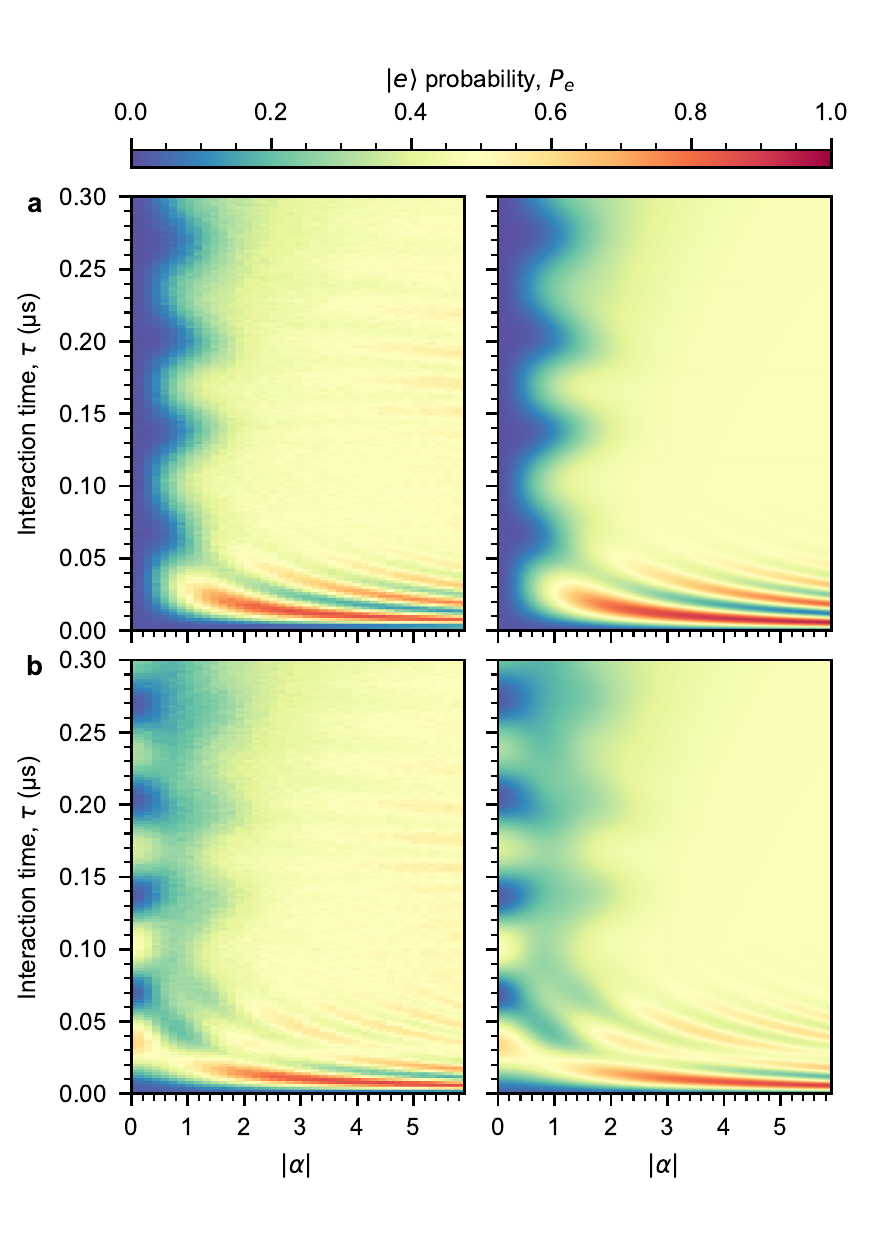}
	\end{center}
	\caption{
		\label{fig:s_alpha}
		{\bf Qubit interaction with larger displaced states.} 
		Left: Experiment. Right: Numerical model. The qubit begins in $|g\rangle$ and interacts with a displaced resonator state.
		{\bf a,} Initial resonator state $D_\alpha|0\rangle=|\alpha\rangle$.
		{\bf b,} Initial resonator state $D_\alpha|1\rangle$.
	}
\end{figure}

We also explore larger displacement amplitudes in addition to those shown in Fig.~\ref{fig:wigner}, as displayed in Fig.~\ref{fig:s_alpha}. These are similar to the Wigner tomography experiments. Fig.~\ref{fig:s_alpha}a has the qubit interact with a coherent state $|\alpha\rangle$; for larger $\alpha$, we see the higher frequencies characteristic of higher harmonic oscillator levels. In Fig.~\ref{fig:s_alpha}b, the resonator is prepared in $|1\rangle$ prior to the displacement. We see excellent agreement between the experiment and model. One interesting feature present in the experiment, but not captured by the model, is the weak revivals around \SI{0.2}{\micro s} and $|\alpha|=5$. This may involve interactions with higher qubit levels; the model only uses two. For this modeling, we use 50 harmonic oscillator levels.

\begin{figure}
	\begin{center}
		\includegraphics[width=3.5in]{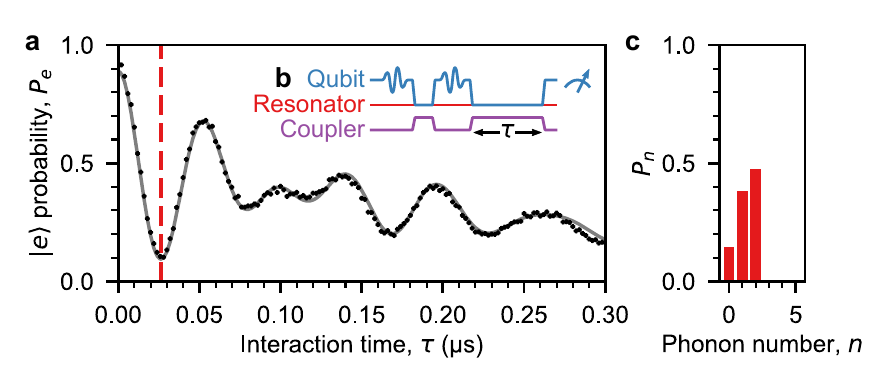}
	\end{center}
	\caption{
		\label{fig:s_two}
		{\bf Generation of the $|2\rangle$ state.} 
		{\bf a,} Qubit evolution nominally starting in $|e,1\rangle$. Black points: Experiment. Gray line: Numerical model. Red dashed line: Time when resonator state is closest to $|2\rangle$.
		{\bf b,} Experiment pulse sequence. The qubit is excited ($X_\pi$), that is swapped into the resonator, the qubit is excited again ($X_\pi$), and then the qubit interacts with the resonator for time $\tau$.
		{\bf c,} The phonon number probability distribution suggested by the model at the red dashed line in {\bf a}.
	}
\end{figure}

We attempt to create the higher Fock state $|2\rangle$ in the SAW resonator by twice exciting the qubit and swapping its excitation into the resonator. We show the result in Fig.~\ref{fig:s_two}. The experiment is limited by the resonator lifetime $T_{1r}$, which is comparable to the duration of the pulse sequence to generate $|2\rangle$, about \SI{100}{ns}. We do observe higher-frequency oscillations in the initial interaction, as expected. The experiment is in excellent agreement with the model, which was fitted to the experiment in the same way as in the Wigner tomography. The resonator state is closest to $|2\rangle$ at the minimum around \SI{26}{ns}. At that time, the resonator state suggested by the model is a statistical mixture of 0.473 $|2\rangle$, 0.382 $|1\rangle$, and 0.145 $|0\rangle$.

\begin{thebibliography}{10}
\expandafter\ifx\csname url\endcsname\relax
  \def\url#1{\texttt{#1}}\fi
\expandafter\ifx\csname urlprefix\endcsname\relax\def\urlprefix{URL }\fi
\providecommand{\bibinfo}[2]{#2}
\providecommand{\eprint}[2][]{\url{#2}}

\bibitem{Stannigel2010}
\bibinfo{author}{Stannigel, K.}, \bibinfo{author}{Rabl, P.},
  \bibinfo{author}{S{\o}rensen, A.~S.}, \bibinfo{author}{Zoller, P.} \&
  \bibinfo{author}{Lukin, M.~D.}
\newblock \bibinfo{title}{Optomechanical transducers for long-distance quantum
  communication}.
\newblock \emph{\bibinfo{journal}{Physical Review Letters}}
  \textbf{\bibinfo{volume}{105}} (\bibinfo{year}{2010}).

\bibitem{Kurizki2015}
\bibinfo{author}{Kurizki, G.} \emph{et~al.}
\newblock \bibinfo{title}{Quantum technologies with hybrid systems}.
\newblock \emph{\bibinfo{journal}{Proceedings of the National Academy of
  Sciences}} \textbf{\bibinfo{volume}{112}}, \bibinfo{pages}{3866--3873}
  (\bibinfo{year}{2015}).

\bibitem{OConnell2010}
\bibinfo{author}{O'Connell, A.~D.} \emph{et~al.}
\newblock \bibinfo{title}{Quantum ground state and single-phonon control of a
  mechanical resonator}.
\newblock \emph{\bibinfo{journal}{Nature}} \textbf{\bibinfo{volume}{464}},
  \bibinfo{pages}{697--703} (\bibinfo{year}{2010}).

\bibitem{Teufel2011}
\bibinfo{author}{Teufel, J.~D.} \emph{et~al.}
\newblock \bibinfo{title}{Sideband cooling of micromechanical motion to the
  quantum ground state}.
\newblock \emph{\bibinfo{journal}{Nature}} \textbf{\bibinfo{volume}{475}},
  \bibinfo{pages}{359--363} (\bibinfo{year}{2011}).

\bibitem{Chan2011}
\bibinfo{author}{Chan, J.} \emph{et~al.}
\newblock \bibinfo{title}{Laser cooling of a nanomechanical oscillator into its
  quantum ground state}.
\newblock \emph{\bibinfo{journal}{Nature}} \textbf{\bibinfo{volume}{478}},
  \bibinfo{pages}{89--92} (\bibinfo{year}{2011}).

\bibitem{Wollman2015}
\bibinfo{author}{Wollman, E.~E.} \emph{et~al.}
\newblock \bibinfo{title}{Quantum squeezing of motion in a mechanical
  resonator}.
\newblock \emph{\bibinfo{journal}{Science}} \textbf{\bibinfo{volume}{349}},
  \bibinfo{pages}{952--955} (\bibinfo{year}{2015}).

\bibitem{Peterson2016}
\bibinfo{author}{Peterson, R.} \emph{et~al.}
\newblock \bibinfo{title}{Laser cooling of a micromechanical membrane to the
  quantum backaction limit}.
\newblock \emph{\bibinfo{journal}{Physical Review Letters}}
  \textbf{\bibinfo{volume}{116}} (\bibinfo{year}{2016}).

\bibitem{Chu2017}
\bibinfo{author}{Chu, Y.} \emph{et~al.}
\newblock \bibinfo{title}{Quantum acoustics with superconducting qubits}.
\newblock \emph{\bibinfo{journal}{Science}} \textbf{\bibinfo{volume}{358}},
  \bibinfo{pages}{199--202} (\bibinfo{year}{2017}).

\bibitem{Morgan2007}
\bibinfo{author}{Morgan, D.}
\newblock \emph{\bibinfo{title}{Surface Acoustic Wave Filters}}
  (\bibinfo{publisher}{Elsevier}, \bibinfo{year}{2007}).

\bibitem{Law1996}
\bibinfo{author}{Law, C.~K.} \& \bibinfo{author}{Eberly, J.~H.}
\newblock \bibinfo{title}{Arbitrary control of a quantum electromagnetic
  field}.
\newblock \emph{\bibinfo{journal}{Physical Review Letters}}
  \textbf{\bibinfo{volume}{76}}, \bibinfo{pages}{1055--1058}
  (\bibinfo{year}{1996}).

\bibitem{Banaszek1999}
\bibinfo{author}{Banaszek, K.}, \bibinfo{author}{Radzewicz, C.},
  \bibinfo{author}{W{\'{o}}dkiewicz, K.} \& \bibinfo{author}{Krasi{\'{n}}ski,
  J.~S.}
\newblock \bibinfo{title}{Direct measurement of the {W}igner function by photon
  counting}.
\newblock \emph{\bibinfo{journal}{Physical Review A}}
  \textbf{\bibinfo{volume}{60}}, \bibinfo{pages}{674--677}
  (\bibinfo{year}{1999}).

\bibitem{Bertet2002}
\bibinfo{author}{Bertet, P.} \emph{et~al.}
\newblock \bibinfo{title}{Direct measurement of the {W}igner function of a
  one-photon fock state in a cavity}.
\newblock \emph{\bibinfo{journal}{Physical Review Letters}}
  \textbf{\bibinfo{volume}{89}} (\bibinfo{year}{2002}).

\bibitem{Haroche2006}
\bibinfo{author}{Haroche, S.} \& \bibinfo{author}{Raimond, J.-M.}
\newblock \emph{\bibinfo{title}{Exploring the quantum: {A}toms, cavities, and
  photons}} (\bibinfo{publisher}{Oxford University Press},
  \bibinfo{year}{2006}).

\bibitem{Hofheinz2009}
\bibinfo{author}{Hofheinz, M.} \emph{et~al.}
\newblock \bibinfo{title}{Synthesizing arbitrary quantum states in a
  superconducting resonator}.
\newblock \emph{\bibinfo{journal}{Nature}} \textbf{\bibinfo{volume}{459}},
  \bibinfo{pages}{546--549} (\bibinfo{year}{2009}).

\bibitem{Vlastakis2013}
\bibinfo{author}{Vlastakis, B.} \emph{et~al.}
\newblock \bibinfo{title}{Deterministically encoding quantum information using
  100-photon {S}chr{\"o}dinger cat states}.
\newblock \emph{\bibinfo{journal}{Science}} \textbf{\bibinfo{volume}{342}},
  \bibinfo{pages}{607--610} (\bibinfo{year}{2013}).

\bibitem{Schuetz2015}
\bibinfo{author}{Schuetz, M.} \emph{et~al.}
\newblock \bibinfo{title}{Universal quantum transducers based on surface
  acoustic waves}.
\newblock \emph{\bibinfo{journal}{Physical Review X}}
  \textbf{\bibinfo{volume}{5}} (\bibinfo{year}{2015}).

\bibitem{Whiteley2018prep}
\bibinfo{author}{Whiteley, S.~J.} \emph{et~al.}
\newblock \bibinfo{title}{Coherent control of spins with {G}aussian acoustics}.
\newblock \emph{\bibinfo{journal}{Submitted}}  (\bibinfo{year}{2018}).

\bibitem{Bohr1920}
\bibinfo{author}{Bohr, N.}
\newblock \bibinfo{title}{{\"U}ber die serienspektra der elemente}.
\newblock \emph{\bibinfo{journal}{Zeitschrift f{\"u}r Physik}}
  \textbf{\bibinfo{volume}{2}}, \bibinfo{pages}{423--469}
  (\bibinfo{year}{1920}).

\bibitem{Gustafsson2014}
\bibinfo{author}{Gustafsson, M.~V.} \emph{et~al.}
\newblock \bibinfo{title}{Propagating phonons coupled to an artificial atom}.
\newblock \emph{\bibinfo{journal}{Science}} \textbf{\bibinfo{volume}{346}},
  \bibinfo{pages}{207--211} (\bibinfo{year}{2014}).

\bibitem{Manenti2017}
\bibinfo{author}{Manenti, R.} \emph{et~al.}
\newblock \bibinfo{title}{Circuit quantum acoustodynamics with surface acoustic
  waves}.
\newblock \emph{\bibinfo{journal}{Nature Communications}}
  \textbf{\bibinfo{volume}{8}} (\bibinfo{year}{2017}).

\bibitem{Arcizet2011}
\bibinfo{author}{Arcizet, O.} \emph{et~al.}
\newblock \bibinfo{title}{A single nitrogen-vacancy defect coupled to a
  nanomechanical oscillator}.
\newblock \emph{\bibinfo{journal}{Nature Physics}}
  \textbf{\bibinfo{volume}{7}}, \bibinfo{pages}{879--883}
  (\bibinfo{year}{2011}).

\bibitem{Kolkowitz2012}
\bibinfo{author}{Kolkowitz, S.} \emph{et~al.}
\newblock \bibinfo{title}{Coherent sensing of a mechanical resonator with a
  single-spin qubit}.
\newblock \emph{\bibinfo{journal}{Science}} \textbf{\bibinfo{volume}{335}},
  \bibinfo{pages}{1603--1606} (\bibinfo{year}{2012}).

\bibitem{Yeo2013}
\bibinfo{author}{Yeo, I.} \emph{et~al.}
\newblock \bibinfo{title}{Strain-mediated coupling in a quantum
  dot{\textendash}mechanical oscillator hybrid system}.
\newblock \emph{\bibinfo{journal}{Nature Nanotechnology}}
  \textbf{\bibinfo{volume}{9}}, \bibinfo{pages}{106--110}
  (\bibinfo{year}{2013}).

\bibitem{Lee2016}
\bibinfo{author}{Lee, K.~W.} \emph{et~al.}
\newblock \bibinfo{title}{Strain coupling of a mechanical resonator to a single
  quantum emitter in diamond}.
\newblock \emph{\bibinfo{journal}{Physical Review Applied}}
  \textbf{\bibinfo{volume}{6}} (\bibinfo{year}{2016}).

\bibitem{Cleland2004}
\bibinfo{author}{Cleland, A.~N.} \& \bibinfo{author}{Geller, M.~R.}
\newblock \bibinfo{title}{Superconducting qubit storage and entanglement with
  nanomechanical resonators}.
\newblock \emph{\bibinfo{journal}{Physical Review Letters}}
  \textbf{\bibinfo{volume}{93}} (\bibinfo{year}{2004}).

\bibitem{Chu2018prep}
\bibinfo{author}{Chu, Y.} \emph{et~al.}
\newblock \bibinfo{title}{Climbing the phonon {F}ock state ladder}.
\newblock \emph{\bibinfo{journal}{In Preparation}}  (\bibinfo{year}{2018}).

\bibitem{Koch2007}
\bibinfo{author}{Koch, J.} \emph{et~al.}
\newblock \bibinfo{title}{Charge-insensitive qubit design derived from the
  {C}ooper pair box}.
\newblock \emph{\bibinfo{journal}{Physical Review A}}
  \textbf{\bibinfo{volume}{76}} (\bibinfo{year}{2007}).

\bibitem{Barends2013}
\bibinfo{author}{Barends, R.} \emph{et~al.}
\newblock \bibinfo{title}{Coherent {J}osephson qubit suitable for scalable
  quantum integrated circuits}.
\newblock \emph{\bibinfo{journal}{Physical Review Letters}}
  \textbf{\bibinfo{volume}{111}} (\bibinfo{year}{2013}).

\bibitem{Chen2014}
\bibinfo{author}{Chen, Y.} \emph{et~al.}
\newblock \bibinfo{title}{Qubit architecture with high coherence and fast
  tunable coupling}.
\newblock \emph{\bibinfo{journal}{Physical Review Letters}}
  \textbf{\bibinfo{volume}{113}} (\bibinfo{year}{2014}).

\bibitem{Geerlings2013}
\bibinfo{author}{Geerlings, K.} \emph{et~al.}
\newblock \bibinfo{title}{Demonstrating a driven reset protocol for a
  superconducting qubit}.
\newblock \emph{\bibinfo{journal}{Physical Review Letters}}
  \textbf{\bibinfo{volume}{110}} (\bibinfo{year}{2013}).

\bibitem{Kelly2015thesis}
\bibinfo{author}{Kelly, J.~S.}
\newblock \emph{\bibinfo{title}{Fault-tolerant superconducting qubits}}.
\newblock Ph.D. thesis, \bibinfo{school}{University of California Santa
  Barbara} (\bibinfo{year}{2015}).

\bibitem{Dunsworth2017}
\bibinfo{author}{Dunsworth, A.} \emph{et~al.}
\newblock \bibinfo{title}{Characterization and reduction of capacitive loss
  induced by sub-micron {J}osephson junction fabrication in superconducting
  qubits}.
\newblock \emph{\bibinfo{journal}{Applied Physics Letters}}
  \textbf{\bibinfo{volume}{111}}, \bibinfo{pages}{022601}
  (\bibinfo{year}{2017}).

\bibitem{Neill2016}
\bibinfo{author}{Neill, C.} \emph{et~al.}
\newblock \bibinfo{title}{Ergodic dynamics and thermalization in an isolated
  quantum system}.
\newblock \emph{\bibinfo{journal}{Nature Physics}}
  \textbf{\bibinfo{volume}{12}}, \bibinfo{pages}{1037--1041}
  (\bibinfo{year}{2016}).

\bibitem{Geller2015}
\bibinfo{author}{Geller, M.~R.} \emph{et~al.}
\newblock \bibinfo{title}{Tunable coupler for superconducting {X}mon qubits:
  Perturbative nonlinear model}.
\newblock \emph{\bibinfo{journal}{Physical Review A}}
  \textbf{\bibinfo{volume}{92}} (\bibinfo{year}{2015}).

\bibitem{Jeffrey2014}
\bibinfo{author}{Jeffrey, E.} \emph{et~al.}
\newblock \bibinfo{title}{Fast accurate state measurement with superconducting
  qubits}.
\newblock \emph{\bibinfo{journal}{Physical Review Letters}}
  \textbf{\bibinfo{volume}{112}} (\bibinfo{year}{2014}).

\bibitem{Kelly2015}
\bibinfo{author}{Kelly, J.} \emph{et~al.}
\newblock \bibinfo{title}{State preservation by repetitive error detection in a
  superconducting quantum circuit}.
\newblock \emph{\bibinfo{journal}{Nature}} \textbf{\bibinfo{volume}{519}},
  \bibinfo{pages}{66--69} (\bibinfo{year}{2015}).

\bibitem{Ebata1988}
\bibinfo{author}{Ebata, Y.}
\newblock \bibinfo{title}{Suppression of bulk-scattering loss in {SAW}
  resonator with quasi-constant acoustic reflection periodicity}.
\newblock In \emph{\bibinfo{booktitle}{{IEEE} Ultrasonics Symposium
  Proceedings.}} (\bibinfo{publisher}{{IEEE}}, \bibinfo{year}{1988}).

\bibitem{Hashimoto2004}
\bibinfo{author}{Hashimoto, K.}, \bibinfo{author}{Omori, T.} \&
  \bibinfo{author}{Yamaguchi, M.}
\newblock \bibinfo{title}{Design considerations on surface acoustic wave
  resonators with significant internal reflection in interdigital transducers}.
\newblock \emph{\bibinfo{journal}{{IEEE} Transactions on Ultrasonics,
  Ferroelectrics and Frequency Control}} \textbf{\bibinfo{volume}{51}},
  \bibinfo{pages}{1394--1403} (\bibinfo{year}{2004}).

\bibitem{Johansson2012}
\bibinfo{author}{Johansson, J.}, \bibinfo{author}{Nation, P.} \&
  \bibinfo{author}{Nori, F.}
\newblock \bibinfo{title}{{QuTiP}: An open-source python framework for the
  dynamics of open quantum systems}.
\newblock \emph{\bibinfo{journal}{Computer Physics Communications}}
  \textbf{\bibinfo{volume}{183}}, \bibinfo{pages}{1760--1772}
  (\bibinfo{year}{2012}).

\bibitem{bertlmann2008}
\bibinfo{author}{Bertlmann, R.~A.} \& \bibinfo{author}{Krammer, P.}
\newblock \bibinfo{title}{Bloch vectors for qudits}.
\newblock \emph{\bibinfo{journal}{Journal of Physics A: Mathematical and
  Theoretical}} \textbf{\bibinfo{volume}{41}}, \bibinfo{pages}{235303}
  (\bibinfo{year}{2008}).

\end{thebibliography}
\end{document}